# Knowledge Recognition Algorithm enables P = NP

Han Xiao Wen
PKU Biocity No. 39 Shang Di Xi Lu, Haidian
Beijing, 100085 China

Summary

This paper introduces a knowledge recognition algorithm (KRA) that is both a Turing machine algorithm and an Oracle Turing machine algorithm. By definition KRA is a non-deterministic language recognition algorithm. Simultaneously it can be implemented as a deterministic Turing machine algorithm. KRA applies mirrored perceptual-conceptual languages to learn member-class relations between the two languages iteratively and retrieve information through deductive and reductive recognition from one language to another. The novelty of KRA is that the conventional concept of relation R $\subseteq$ {$\Sigma^*$ x $\Sigma^*$} is adjusted to R $\subseteq$ {$\Sigma^*_p = \Sigma^*_c \cup \Sigma^*_{|p|} \in \Sigma^*_{|c|}$}. The computation therefore becomes efficient bidirectional string mapping.

## 1. Descriptions of Knowledge Recognition Algorithm

Knowledge recognition (also called relation recognition) algorithm (KRA), was originally designed to simulate the mirrored language structure of the human brain by Han (Han08). The human brain contains a mirrored perceptual-conceptual language structure for storing member-class relations between the two languages as knowledge. That is, KRA has two levels of languages, which permit the perceptual language $L_p$ as the members of the conceptual language $L_c$, and conceptual language as the class of the perceptual. Based on this continuous iterative structure, four "innate" logic functions exist, defined by four axioms:

**Sensation**: Innate mapping function of *one-to-one correspondence* $L_p \supseteq p = c \in L_c$ exists between the perceptual language $L_p$ and conceptual language $L_c$. The existence of sensation also can be presented equivalently as the "diagonal" set {$(p,c) \mid p = c$} (see Fig1).

**Induction**: *Learning* function of *member-class relation* $L_p \supseteq |p| \in |c| \in L_c$ exists between the perceptual language $L_p$ and conceptual language $L_c$, where $|p|$ and $|c|$ denote the length level of $p$ and $c$, respectively. The existence of induction also can be presented equivalently as the "membership" set {$(p,c) \mid |p| \in |c|$ }.

**Deduction**: *Class recognition* function exists for recognizing relations from the perceptual language $L_p$ to conceptual language $L_c$, defined as follows: Suppose that $L_k$ is a language of knowledge over $\Sigma_k$, k = p, c, Then $L_p \geq L_c$ iff $p = c$ and $|p| \in |c|$ such that $p \in \Sigma^*_p \rightarrow c \in \Sigma^*_c$, for all $p \in \Sigma^*_p$ and $c \in \Sigma^*_c$.

**Reduction**: *Membership recognition* function exists for recognizing relations from the conceptual language $L_c$ to perceptual language $L_p$, defined as follows: Suppose that $L_k$ is a language of knowledge over $\Sigma_k$, k = p, c, Then $L_p \leq L_c$ iff $p = c$ and $|p| \in |c|$ such that $p \in \Sigma_p^*$ ← $c \in \Sigma_c^*$, for all $p \in \Sigma_p^*$ and $c \in \Sigma_c^*$. This function is equivalent to the notion of reducibility. It is an inverse function of deduction.

Formally KRA is a *string mapping* language $L_k$ over $\Sigma_k$, k = p, c. Let $\Sigma_p$ and $\Sigma_c$ be two identical sets over a binary alphabet, and let $\Sigma_p^*$ and $\Sigma_c^*$ be two sets of finite identical strings over $\Sigma_p$ and $\Sigma_c$. Then the *language $L_p$ over $\Sigma_p$* is a subset of $\Sigma_p^*$, and the *language $L_c$ over $\Sigma_c$* is a subset of $\Sigma_c^*$. Thus $L_p$ and $L_c$ are identical languages, denoted by $L_p \supseteq p = c \in L_c$. There exists a binary relation $R \subseteq \Sigma_p^* \times \Sigma_c^*$ for some finite alphabets $\Sigma_p$ and $\Sigma_c$. We associate with each such relation $R$ a language $L_R$ over $\Sigma_p^* \cup \Sigma_c^* \cup \{\#\}$ defined by

$$L_R = \{p \# c \mid R(p, c)\}$$

where the symbol # is not in $\Sigma_p$ and $\Sigma_c$.

There exists a *subset one-to-one correspondence* $R_=^* = \{\Sigma_p^* = \Sigma_c^*\}$ in $R \subseteq \Sigma_p^* \times \Sigma_c^*$.

There also exists a *subset member-class relation* $R_\in^* = \{\Sigma_{|1|}^* \in \Sigma_{|2|}^* \in \Sigma_{|3|}^* \ldots \in \Sigma_{|n-1|}^* \in \Sigma_{|n|}^*\} = \{\Sigma_{|p|}^* \in \Sigma_{|c|}^*\}$ in $R \subseteq \Sigma_p^* \times \Sigma_c^*$, where |1|, |2|, |3|… |n-1|, |n| denote the length of the binary strings in the continued iterative perceptual-conceptual relations.

Then relation $L_R$ is the union of $\Sigma_p^* = \Sigma_c^*$ and $\{\Sigma_{|p|}^* \in \Sigma_{|c|}^*\}$ iteratively, defined by

$$L_R = \{p \# c \mid \Sigma_p^* = \Sigma_c^* \cup \Sigma_{|p|}^* \in \Sigma_{|c|}^*\}$$

The notion *string mapping* is to map member-class relations, where *deduction* is the mapping from perceptual to conceptual language denoted by $L_p \geq L_c$, and *reduction* is the mapping from conceptual to perceptual language denoted by $L_p \leq L_c$.

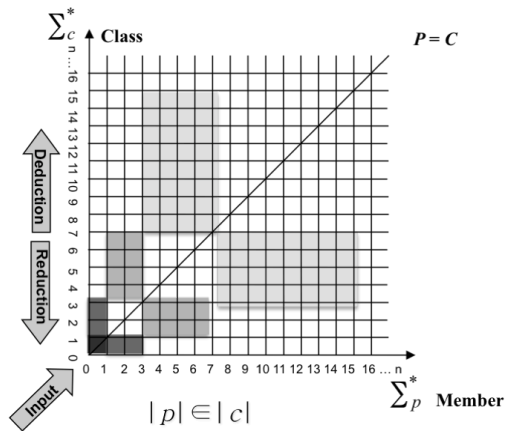

Fig1. Iterations of member-class relations between perceptual and conceptual language

## 2. KRA enables P = NP

We denote by $t_M(w)$ the number of steps in the computation of $M$ on input $w$. We denote by $T_M(n)$ the *worst case run time* of $M$; that is,

$$T_M(n) = \max\{t_M(w) \mid w \in \Sigma_{|1|}^* \in \Sigma_{|2|}^* \in \Sigma_{|3|}^* \ldots \in \Sigma_{|n-1|}^* \in \Sigma_{|n|}^*\}$$

It is easy to see that KRA can answer membership and class relation questions of the form $L_p \ni |p| \in |c| \in L_c$ correctly in polynomial time.

Theorem 1. $L_k$ is both a P and an NP language (P = $L_k$ = NP) iff $L_k$ is over $\Sigma_k$, k = p, c, where $L_p \ni p = c \in L_c$, and $L_p \ni |p| \in |c| \in L_c$, for all $p \in \Sigma_p^*$ and $c \in \Sigma_c^*$.

Proof. $L_k$ is an NP language by definition where there is a countable domain D set, a finite alphabet $\Delta$ such that $\Delta^* \cap \{\text{ACCEPT, REJECT}\} = \phi$, an encoding function E: D $\to \Delta^*$, a transition relation $\tau$: $\subseteq \Delta^* \times (\Delta^* \cup \{\text{ACCEPT, REJECT}\})$, such that $p \in L_p \Leftrightarrow f(p) \in L_c$, for all $p \in \Sigma_p^*$.

$L_k$ is a P language by definition where there is a countable domain set D, a countable range R = D, a finite alphabet $\Delta$ such that $\Delta^* \wedge R = \phi$, an encoding function E: D $\to \Delta^*$, a transition function $\tau$: $\Delta^* \to \Delta^* \cup R$, such that $r(p, c) \Leftrightarrow p \in L_k$, for all $p, c \in \Sigma_k^*$. Thus P = $L_k$ = NP.

P = $L_k$ = NP means that the deterministic algorithm can take efficiency advantage of relation mapping where relations are learned into and retrieved from a polynomial space in polynomial time through relation recognition. The conventional concept of relation R $\subseteq \{\Sigma^* \times \Sigma^*\}$ is adjusted to R $\subseteq \{\Sigma_p^* = \Sigma_c^* \cup \Sigma_{|p|}^* \in \Sigma_{|c|}^*\}$. The computation becomes a pure string mapping such that, for every $y_o \in \Delta^*$, the set $\{<y_o, y> | <y_o, y> \in \tau\}$ has fewer than $k_A$ elements, where $k_A$ is a constant. The computation of A on input $x \in D$ is a sequence $y_1, y_2, \ldots$ which ends with $y_K$, such that $y_1 = E(x)$, $<y_i, y_{i+1}> \in \tau$ for all i, and $y_K \in \{\text{ACCEPT, REJECT}\}$. [Coo00, Kar72]

**Acknowledgement**